\documentclass[aps,prd,twocolumn,superscriptaddress,eqsecnum,nofootinbib,showpacs,preprintnumbers]{revtex4}
\usepackage{graphicx}
\usepackage{euscript,amssymb}
\usepackage{amsfonts}
\usepackage{amsmath}
\usepackage{amssymb}
\usepackage{graphicx}
\usepackage{euscript}

\begin{document}

\title{Escaping the Big Rip?}

\author{Mariam Bouhmadi-L\'{o}pez}
\affiliation{Institute of Cosmology and Gravitation, University of
Portsmouth,  Mercantile House, Hampshire Terrace, Portsmouth PO1
2EG, UK} \email{mariam.bouhmadi@port.ac.uk}

\author{Jos\'{e} A. Jim\'{e}nez Madrid}
\affiliation{Instituto de Astrof\'{\i}sica de Andaluc\'{\i}a
(CSIC), Camino Bajo de Hu\'{e}tor 24, 18008 Granada, Spain
\\  and \\ Instituto de Matem\'{a}ticas y F\'{\i}sica Fundamental
(CSIC), Serrano 121, 28006 Madrid, Spain} \email{madrid@iaa.es}

\pacs{98.80.-k,98.80.Es,11.10.-z \hfill PU-ICG-04/09,
astro-ph/0404540}

\begin{abstract}
We discuss dark energy models which might describe effectively the
actual acceleration of the universe. More precisely, for a
4-dimensional Friedmann-Lema\^{\i}tre-Robertson-Walker (FLRW)
universe we consider two situations: First of them, we model dark
energy by phantom energy described by a perfect fluid satisfying
the equation of state $P=(\beta-1)\rho$ (with $\beta<0$ and
constant). In this case the universe reaches a ``Big Rip''
independently of the spatial geometry of the FLRW universe. In the
second situation, the dark energy is described by a phantom
(generalized) Chaplygin gas which violates the dominant energy
condition. Contrary to the previous case, for this material
content a FLRW universe would never reach a ``big rip''
singularity (indeed, the geometry is asymptotically de Sitter). We
also show how this dark energy model can be described in terms of
scalar fields, corresponding to a minimally coupled scalar field,
a Born-Infeld scalar field and a generalized Born-Infeld scalar
field. Finally, we introduce a phenomenologically viable model
where dark energy is described by a phantom generalized Chaplygin
gas.
\end{abstract}

\date{April 27, 2004}
\maketitle

\section{Introduction}

Several astronomical  and cosmological observations,  ranging from
the cosmic microwave background anisotropy \cite{{Spergel:2003cb}}
to observation of distant supernova \cite{Riess:1998cb}, show that
the universe is undergoing an accelerating stage. In addition,
these observations show that the acceleration of the universe is
due to some unknown stuff  usually dubbed dark energy (DE), which
constitutes roughly two thirds of the total energy density of the
universe. Moreover, it is known that the DE satisfies an equation
of state $P=(\beta-1)\rho$, where $|\beta|<0.3$ (at least recently
in the history of the universe) \cite{Jimenez:2003nz}.

So far, several phenomenological models have been proposed to
describe the dark energy,  being the cosmological constant,
$\Lambda$,  by far the most simple and popular candidate
\cite{Lambda}. However, this possibility is  ruled out (in
principle) as a consequence of the huge discrepancy between the
expected theoretical and experimental value of $\Lambda$. A
positive cosmological constant might describe the acceleration of
the universe as it could be described as a perfect fluid with
negative pressure, $-\Lambda$,  and this is one of the main
ingredients to produce an accelerating universe. In fact, a
Friedmann-Lema\^{\i}tre-Robertson-Walker (FLRW) universe undergoes
an accelerated stage as long as $\rho+3P<0$, where $\rho$ and $P$
correspond, respectively, to the total energy density and pressure
of the matter content. Matter contents with this requirement can
be described effectively, for example, by a perfect fluid; with a
barotropic equation of state or a more exotic one like in
(generalized) Chaplygin gas models
\cite{chaplygin,chaplygin2,MBL-PVM}, or by dynamical scalar fields
as in quintessence models \cite{quintessence} and  phantom energy
models \footnote{We would like to mention that there are other
candidates for dark energy based on brane-world models
\cite{brane} and modified 4-dimensional Einstein-Hilbert
actions\cite{EHmodified}, where a late time acceleration of the
universe may be achieved.}
\cite{caldwell,SSD,phantom1,Li,phantom2,phantom3}.

If  dark energy would be  described by either of the last two
models mentioned above, then  the future of the universe might be
quite different. While for a quintessence scalar field, with an
effective equation of state $P=(\beta-1)\rho$, with $\beta$
constant and \mbox{$0<\beta<{2}/{3}$}, dominating the energy
density of the universe, the universe would expand forever, for a
phantom energy; i.e. $\beta<0$,  this might not be the case. In
fact, for a matter content with a barotropic equation of state
formally similar to the previous one, but with a negative $\beta$,
the universe would experiment a cosmic doomsday, also dubbed  big
rip, \cite{phantom1,caldwell,SSD,Li,phantom2,phantom3}; i.e. the
scale factor would blow up in a finite cosmic time. The last
affirmation is based on a constant negative value of $\beta$.
However, the value of $\beta$ may change along the evolution of
the universe and then in principle the universe might not reach a
big rip in the future.

Another candidate to describe dark energy is a (generalized)
Chaplygin gas \cite{chaplygin}, already mentioned, which
corresponds to a perfect fluid with a rather strange equation of
state $P=-A/\rho^\alpha$, where $A$ is a positive constant and
$\alpha$ a parameter. This fluid can describe a transition from a
dust dominated universe at early time to a de Sitter universe at
late time. In addition, this matter content has been proposed as a
unification of dark matter and dark energy. In this paper, we will
show that if the dark energy is modelled by a phantom generalized
Chaplygin gas, then the universe will escape the big rip and will
expand forever. Others phantom energy models; i.e. matter contents
with $\rho+P<0$ and a positive energy density, exhibiting  similar
property  has already been proposed \cite{SSD,Li}. For example,
this can be achieved considering a homogeneous minimally coupled
scalar field with an appropriate potential \cite{SSD} or with a
Born-Infeld homogeneous scalar field \cite{Li}. However, in these
models, the scalar field has the wrong kinetic energy. In this
paper, we propose an alternative phenomenological model to
describe phantom energy by means of a fluid which, firstly,
satisfies the generalized Chaplygin gas equation of state
\cite{chaplygin}, and secondly, violates the dominant energy
condition \cite{EllisHawking}. Moreover, for this peculiar
material content a FLRW universe would never reach a Big Rip.

The paper can be outlined as follows. In the next section, on the
one hand, we will review the phantom energy model with a constant
equation of state ($\beta$ is negative and constant) in FLRW
universes, giving the explicit expression of the scale factor for
the three different spatial geometries. On the other hand, we will
discuss if the presence of a positive cosmological constant might
alleviate the big rip problem. In section \ref{3}, we introduce
and study a dark energy model based on a generalized Chaplygin
gas. In section \ref{4}, we analyze our model in the light of
scalar fields, corresponding to a minimally coupled scalar field,
a Born-Infeld scalar field and a generalized Born-Infeld scalar
field. In section \ref{viable}, we describe a phenomenologically
viable model, where the dark energy is described by a phantom
generalized Chaplygin gas. Finally, we briefly summarize and
discuss our results in section \ref{conclusion}.

\section{Phantom energy \label{2}}

Through the paper, we  mainly consider the late time evolution of
a homogeneous and isotropic universe.  Furthermore, we model dark
energy by  phantom energy, which in the present section is
described by a perfect fluid satisfying the equation of state
$P=(\beta-1)\rho$, where $\beta$ is constant and negative. The
conservation equation results on $\rho=\tilde A a^{-3\beta}$,
where $\tilde A$ is an integration constant. Therefore, for
$\beta<0$ the energy density grows with the scale factor instead
of decreasing. For simplicity, we disregard the other matter
contents of the universe as their energy densities decrease with
the cosmic time and can be neglected in comparison with the energy
density of phantom matter at very late time when a big rip could
happen \footnote{In section \ref{viable}, we consider the other
matter components of the universe together with a phantom matter,
defined in section \ref{3}, and constraints the model using the
observational cosmological parameters.}. Consequently, the
Friedmann equation can be expressed as
\begin{equation}
H^2+\frac{k}{a^2}=\frac{8\pi\textrm{G}}{3}\tilde A a^{-3\beta},
\label{phantomFriedmann}
\end{equation}
where $H$ is the Hubble constant and $k=1,-1,0$, corresponding to
spherical, hyperbolic or flat spatial sections of the FLRW model.

\begin{figure}[h]
\includegraphics[width=0.8\columnwidth]{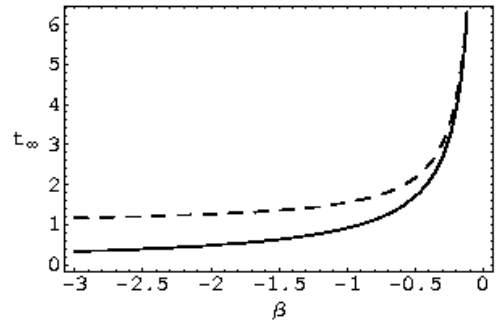}
\caption{This figure shows the behavior of the cosmic time
corresponding to the big rip as a function  of the parameter
$\beta$ related to the ratio between the pressure and the energy
density of the phantom matter. The dashed line corresponds to the
case of a FLRW universe sliced into hyperbolic sections. The solid
line corresponds to a homogeneous and isotropic space-time sliced
into spherical sections. The cosmic time has been divided by
$C^{1/(3\beta-2)}$.} \label{bigrip}
\end{figure}

For flat spatial section ($k=0$) the scale factor scales with the
cosmic time, t,  as
\begin{equation}
a(t)=\left[a_0^{3\beta/2}+\frac{3\beta}{2}C^{1/2}(t-t_0)\right]^{2/(3\beta)},
\label{phantomflat}
\end{equation}
where $a_0$ and $t_0$ will be integration constants throughout the
paper, corresponding to the initial radius and cosmic time of the
universe\footnote{The integration constant $t_0$ can be set equal
to zero. However, $a_0$ must be different from zero, otherwise the
scale factor will be vanishing at any cosmic time.} and
\mbox{$C=(8\pi\textrm{G}/3)\tilde A$.} As can be seen, for
negative $\beta$, the scale factor diverges in a finite cosmic
time
\begin{equation}
t_{\infty 0}=t_0-\frac{2}{3\beta C^{1/2}}a_0^{3\beta/2},
\label{tbigrip+0}
\end{equation}
if $t$ varies between $t_0$ and $t_{\infty 0}$. Therefore, the
universe would reach a cosmic doomsday \cite{caldwell}. For values
of the cosmic time larger than $t_{\infty 0}$ the scale factor
will decrease until vanishing at $t \rightarrow +\infty$. It can
also be seen that when $\beta$ approaches zero, $t_{\infty 0}$
goes to infinity. A similar situation can be found in the case of
a FLRW universe with hyperbolic or spherical spatial geometry [see
Fig.~\ref{bigrip}]. This is not surprising as for a vanishing
$\beta$, the FLRW geometry corresponds to a de Sitter space-time
sliced into flat sections and there is no longer a big rip. In
addition, for a given initial value of the  scale factor $a_0
>\exp[2/(3\beta)]$, the larger is the value of $\beta$, the
larger is the value of $t_{\infty 0}$ at which the big rip
happens. Moreover, the latter the phantom energy starts dominating
the energy density of the universe; i.e. the larger the value of
$a_0$, the sooner the universe reaches the big rip.

In the spherical case ($k=1$), the scale factor must be larger
than $a_{\textrm{min}}=C^{1/(3\beta-2)}$. Otherwise, the Friedmann
equation (\ref{phantomFriedmann}) is not well defined. We have
that the cosmic time scales with the scale factor as \cite{W}
\begin{eqnarray}
t-t_0&=&\frac{2C^{1/(3\beta-2)}}{2-3\beta}\;\left({C
a^{2-3\beta}-1}\right)^{1/2}\nonumber\\
&\times&\textrm{F}\left(
\frac{3\beta-1}{3\beta-2},\frac12;\frac32;1-Ca^{2-3\beta}\right),
\label{aphantomspheric1}
\end{eqnarray}
where $\textrm{F}(b,c;d;e)$ is a hypergeometric series \cite{W} .
The cosmic time is finite whenever\footnote{A hypergeometric
series $\textrm{F}(b,c;d;e)$, also called a hypergeometric
function, converges at any value $e$ such that $|e|\leq 1$,
whenever $b+c-d<0$. However, if  $0 \leq  b+c-d < 1$ the series
does not converge at $e=1$. In addition, if  $1 \leq b+c-d$, the
hypergeometric function blows up at $|e|=1$
\cite{W}.\label{series}} $-1\leq 1-Ca^{2-3\beta}\leq 0$, i.e.
\mbox{$a_{\textrm{min}}\leq a\leq
\left(2/C\right)^{1/(2-3\beta)}$}. For larger values of the scale
factor the expression (\ref{aphantomspheric1}) breaks down and,
consequently, we cannot immediately conclude either the existence
or the absence of a cosmic doomsday. However, the difference
between the cosmic times corresponding, respectively, to
$t_{\infty+}$ when the scale factor blows up, and to a given
cosmic time $t$ such that $a(t)$ is larger than
$\left(2/C\right)^{1/(2-3\beta)}$ can be expressed as follows
\begin{eqnarray}
&&t_{\infty+}-t= -\frac{2 C^{1/(3\beta-2)}}{3\beta}\; \left({C
a^{2-3\beta}-1}\right)^{-\frac{3\beta}{2(3\beta-2)}}\nonumber\\
&\times&\textrm{F}\left(\frac{3\beta-1}{3\beta-2},\frac{3\beta}{2(3\beta-2)};
\frac{3\beta}{2(3\beta-2)}+1;\frac{1}{1-Ca^{2-3\beta}}\right).\nonumber\\
\label{aphantomspheric2}
\end{eqnarray}
In addition, it can be checked that the last expression is well
defined, in particular the hypergeometric function (see footnote
\ref{series}), whenever $a$ is larger or equal than
$\left(2/C\right)^{1/(2-3\beta)}$. This value of the scale factor
corresponds precisely to the maximum value allowed in
Eq.~(\ref{aphantomspheric1}). Consequently, we can conclude that
there is a big rip in a FLRW universe sliced into spherical
sections filled with phantom matter when $\beta$ is constant and
negative. In this case, we have that the cosmic time elapsed since
the scale factor acquires its minimum value $a_{\textrm{min}}$ at
$t=t_0$ up to the divergence of the radius of the universe is
\begin{eqnarray}
t_{\infty+}&=&C^{1/(3\beta-2)}\left[\frac{2}{2-3\beta}\;\textrm{F}\left(
\frac{3\beta-1}{3\beta-2},\frac12;\frac32;-1\right)\right.\nonumber\\
&-&\left.\frac{2}{3\beta}\;\textrm{F}\left(\frac{3\beta-1}{3\beta-2},\frac{3\beta}{2(3\beta-2)};
\frac{3\beta}{2(3\beta-2)}+1;-1\right)\right].\nonumber\\
\label{tbigrip+}
\end{eqnarray}

Similarly, a universe filled with phantom matter with $\beta$
constant reaches a big rip in the future,  if the geometry
corresponds to a FLRW universe sliced into hyperbolic sections
($k=-1$). In fact, on the one hand, we have that the cosmic time
varies with the scale factor as
\begin{eqnarray}
t-t_0=a\;\textrm{F}\left(\frac12,-\frac{1}{3\beta-2};-\frac{1}{3\beta-2}+1;-Ca^{2-3\beta}\right),\nonumber\\
\label{aphantomhyperbolic1}
\end{eqnarray}
for $a\leq C^{1/(3\beta-2)}$. On the other hand, we have that for
larger value of the scale factor, the cosmic time reads
\begin{eqnarray}
t_{\infty-}-t&=& -\frac{2}{3\beta{C}^{1/2}}\;
a^{3\beta/2}\nonumber\\
&\times&\textrm{F}\left(\frac12,\frac{3\beta}{2(3\beta-2)};
\frac{3\beta}{2(3\beta-2)}+1;-\frac{1}{C}a^{3\beta-2}\right),\nonumber\\
\label{aphantomhyperbolic2}
\end{eqnarray}
where $t_{\infty-}$ corresponds to the cosmic time when the scale
factor reaches infinite values. The last two expressions are well
defined at $a=C^{1/(3\beta-2)}$ (see footnote \ref{series}). In
this model, it can be seen that the scale factor varies between
zero and infinity in a finite cosmic time corresponding to
\begin{eqnarray}
t_{\infty-}&=&C^{1/(3\beta-2)}\left[\textrm{F}\left(
\frac12,-\frac{1}{3\beta-2};-\frac{1}{3\beta-2}+1;-1\right)\right.\nonumber\\
&-&\left.\frac{2}{3\beta}\;\textrm{F}\left(\frac12,\frac{3\beta}{2(3\beta-2)};
\frac{3\beta}{2(3\beta-2)}+1;-1\right)\right].\nonumber\\
\label{tbigrip-}
\end{eqnarray}

\begin{figure}[h]
\includegraphics[width=0.8\columnwidth]{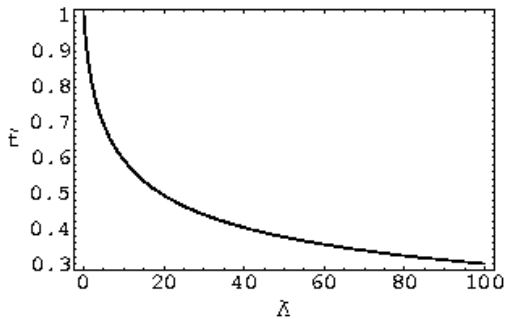}
\caption{This figure shows the behavior of the cosmic time
corresponding to the big rip as a function  of the cosmological
constant. As the graphic shows, the largest value for $\tilde t$
is achieved in the absence of a positive cosmological constant.
Indeed, the larger is $\tilde\Lambda$, the sooner the big rip
takes place. The cosmological constant and the cosmic time in the
graphic are redefined as dimensionless quantities given by
$\tilde\Lambda/(Ca_0^{-3\beta})$ and
\mbox{$-(3/2)\beta\sqrt{C}a_0^{-3\beta/2}(\tilde t-t_0)$,}
respectively.} \label{Lambda}
\end{figure}

Before concluding this section, we will analyze if the inclusion
of a constant positive vacuum energy density; i.e. a positive
cosmological constant $\Lambda$,  may alleviate the big rip
problem dues to  phantom energy with constant equation of state
($\beta$ constant). For simplicity and analyticity, we will
restrict to the case of a FLRW universe with flat spatial
geometry. The  Friedmann equation reads
\begin{equation}
H^2=\frac{\Lambda}{3}+Ca^{-3\beta}.
\end{equation}
The solution to the last equation is
\begin{eqnarray}
a^{3\beta}(t)&=&a_0^{3\beta}(1-D)^{-2}
\left(\exp\left[\frac{3\beta}{2}\sqrt{\tilde\Lambda}(t-t_0)\right]\right.\nonumber\\
&& -\left.
D\exp\left[-\frac{3\beta}{2}\sqrt{\tilde\Lambda}(t-t_0))\right]\right)^2,
\label{phantomlambda}
\end{eqnarray}
where $\tilde\Lambda=\Lambda/3$ and $D$ is a positive constant
given by
\begin{equation}
D=\frac{\sqrt{\tilde\Lambda+Ca_0^{-3\beta}}-
\sqrt{\tilde\Lambda}}{\sqrt{\tilde\Lambda+Ca_0^{-3\beta}}+
\sqrt{\tilde\Lambda}}.
\end{equation}
From Eq.~(\ref{phantomlambda}), it can be seen that the scale
factor grows from an initial value $a_0$ and blows up in a finite
cosmic time; i.e. the universe will face a cosmic doomsday, when
$t$ approaches $\tilde t= t_0+(\ln D)
/(3\beta\sqrt{\tilde\Lambda})$. For $\tilde t < t$, the scale
factor decreases and the universe collapses when $t$ approaches
infinite values. In addition, the larger is the value of $a_0$,
the smaller is $\tilde t$, and consequently, the sooner the cosmic
doomsday happens. A similar conclusion holds for $¦\beta¦$ (at
least for $a_0>1$). Moreover, it can be checked that $\tilde t$
approaches $t_{\infty0}$, defined in Eq.~(\ref{tbigrip+0}), when
the cosmological constant vanishes. In summary, we have that the
presence of a cosmological constant does not modify the general
features of the model and the big rip cannot be avoided for
$\beta$ constant and negative. Moreover, a positive vacuum energy
density cannot delay the happening of the big rip [see
Fig.~\ref{Lambda}]. This can be explained as follows: the presence
of a positive cosmological constant in the model induces a bigger
growth of the Hubble parameter and, consequently, the scale factor
increases faster leading to a sooner big rip.

\section{Generalized Chaplygin gas and phantom energy \label{3}}

The generalized Chaplygin gas can be described as a perfect fluid
with the following equation of state \cite{chaplygin}
\begin{equation}
P=-A/\rho^\alpha, \label{gcg}
\end{equation}
where $A$ is a positive constant and $\alpha$ is a parameter. In
the particular case $\alpha=1$, the equation of state (\ref{gcg})
corresponds to a Chaplygin gas. The conservation of the energy
momentum tensor implies
\begin{equation}
\rho=\left[A+\frac{(\rho_0^{\alpha+1}-A)a_0^{3(\alpha+1)}}{a^{3(\alpha+1)}}
\right]^{\frac{1}{1+\alpha}},
\label{chaplygindensity}\end{equation}
where $a_0$ and $\rho_0$ are the initial scale factor and energy
density, respectively. It can be checked that the dominant energy
condition is fulfilled whenever $A<\rho^{(\alpha+1)}$. This
requirement is strongly related to the initial values of the model
and the specific equation of state through the constant
\begin{equation}
B\equiv (\rho_0^{\alpha+1}-A)a_0^{3(\alpha+1)}. \label{defB}
\end{equation}
For positive values of $B$, $P+\rho$ is positive and the dominant
energy condition is satisfied. This is not the case, when $B$ is
negative. Let us see the behaviour of the energy density for both
cases.

When the parameter $B$ is positive, $\rho$ will be a decreasing
function of $a$. In fact, for $1+\alpha>0$ the generalized
Chaplygin gas interpolates between dust for small scale factors
and a constant energy density at large scale factors. This
property has promoted the generalized Chaplygin gas to be a
candidate to unify dark energy and dark matter \cite{chaplygin}.
For $1+\alpha<0$, the energy density behaves on the other way
round; i.e. $\rho$ approaches $A^{1/(1+\alpha)}$ for small scale
factor and behaves as a pressureless fluid at late time.

When the parameter $B$ is negative, the energy density will be an
increasing function of the scale factor. Moreover, $\rho$ is
larger than $A^{1/(1+\alpha)}$ when $1+\alpha<0$, reaching its
minimum value at $a=0$ and blowing up when the scale factor
approaches its maximum value
\begin{equation}
\bar{a}\equiv\left(-\frac{B}{A}\right)^{1/[3(1+\alpha)]}.
\label{bara}
\end{equation}
In what follows, we will disregard this set up ($B<0$ and
$1+\alpha<0$). On the other hand, if $1+\alpha>0$ and $B<0$, the
scale factor is larger than $\bar{a}$, in such a way that $\rho$
vanishes at this scale factor and approaches $A^{1/(1+\alpha)}$
when the scale factor goes to infinity.  We will henceforth
analyze this last case, which can be included in the set of
phantom energy models as $\rho>0$ and $P+\rho<0$.

As in the previous section, we consider a homogeneous and
isotropic universe, where now the phantom energy is given by a
generalized Chaplygin gas such that $B<0$ and $-1<\alpha$. The
Friedmann equation reads
\begin{equation}
H^2+\frac{k}{a^2}=\frac{8\pi\textrm{G}}{3}\left[A+\frac{B}{a^{3(\alpha+1)}}
\right]^{\frac{1}{1+\alpha}}. \label{gcgFriedmann}
\end{equation}

If the FLRW universe is sliced into flat sections, then the cosmic
time is related to the scale factor as
\begin{eqnarray}
t-t_0&=&\frac{2D^{-\frac12}A^{-\frac{1}{2(1+\alpha)}}}{3(1+2\alpha)}
\left[1+\frac{B}{A}a^{-3(1+\alpha)}\right]^{\frac{1+2\alpha}{2(1+\alpha)}}\nonumber\\
&\times&\textrm{F}\left(1,\frac{1+2\alpha}{2(1+\alpha)};\frac{3+4\alpha}{2(1+\alpha)};1+\frac{B}{A}a^{-3(1+\alpha)}\right),
\nonumber\\
\label{agcg}
\end{eqnarray}
where $D=8\pi G/3$ and $-1/2<\alpha$. Firstly, we have that the
scale factor is bigger than $\bar a$ defined in Eq.~(\ref{bara}).
In this case there is no big rip: when the scale factor blows up,
the cosmic time does too (see footnote \ref{series}). In
opposition with the cases studied in the previous section, the
present model does not show a cosmic doomsday because the Hubble
parameter approaches a constant non vanishing value for large
scale factors. Consequently, at late time the geometry of the
model is asymptotically de Sitter. Although we have not been able
to get an equivalent analytical expression to Eq.~(\ref{agcg}) for
$-1<\alpha<-1/2$, a similar conclusion holds because $H^2$
approaches a positive non vanishing value when $a\rightarrow
+\infty$.

When the spatial geometry of the homogeneous and isotropic
space-time is spherical, the Hubble parameter is well defined as
long as $a>a_{\textrm{min}}$, where $a_{\textrm{min}}$ is such
that $H(a_{\textrm{min}})=0$. The explicit expression of
$a_{\textrm{min}}$ is given in the appendix. It can be shown that
$a_{\textrm{min}}$ is larger than the minimum value of the scale
factor for flat spatial geometry given in Eq.~(\ref{bara}).
Moreover, the cosmic time for $k=1$ satisfies the inequality
\begin{eqnarray}
t-t_0&>&\int_{\bar a}^a \left[D
a^2\left(A+\frac{B}{a^{3(1+\alpha)}}
\right)^{\frac{1}{1+\alpha}}\right]^{-\frac12}da\nonumber\\
&-& \int_{\bar a}^{a_{\textrm{min}}}\left[D
a^2\left(A+\frac{B}{a^{3(1+\alpha)}}
\right)^{\frac{1}{1+\alpha}}\right]^{-\frac12}da.\nonumber\\
\label{inequality+1}
\end{eqnarray}
The second term on the right hand side (rhs) of the inequality is
finite. Indeed, it is the cosmic time for flat spatial sections
corresponding to  $a=a_{\textrm{min}}$. In addition, the first
term on the rhs corresponds  to the cosmic time for $k=0$ geometry
at a given scale factor $a>\bar a$. As can be seen for large scale
factor the cosmic time for $k=1$ blows up because $t-t_0$ diverges
for flat spatial geometry (first term on rhs). Consequently, the
universe does not hit a cosmic doomsday in its future.

Similarly, the cosmic time for a FLRW universe with spatial
hyperbolic sections can be bounded from below as follows
\begin{eqnarray}
t-t_0&>&\frac{1}{\sqrt{2}}\Big\{
a_{\textrm{min}}-\bar{a}\Big.\nonumber\\
&+&\Big. \int_{\bar a}^a \left[D
a^2\left(A+\frac{B}{a^{3(1+\alpha)}}
\right)^{\frac{1}{1+\alpha}}\right]^{-\frac12}da\Big.\nonumber\\
&-& \Big.\int_{\bar a}^{a_{\textrm{min}}}\left[D
a^2\left(A+\frac{B}{a^{3(1+\alpha)}}
\right)^{\frac{1}{1+\alpha}}\right]^{-\frac12}da\Big\},\nonumber\\
\label{inequality-1}\end{eqnarray}
when the matter content corresponds to a generalized Chaplygin
gas. We would like to point out that the last two terms on rhs of
the inequality coincides precisely with the ones on the rhs of the
expression (\ref{inequality+1}). Consequently, based on an
argument similar to that is used for the $k=1$ case, we can
conclude that there is no big rip for $k=-1$. In addition, the
scale factor grows from $\bar a$ to infinity.

\begin{figure}[h]
\includegraphics[width=0.8\columnwidth]{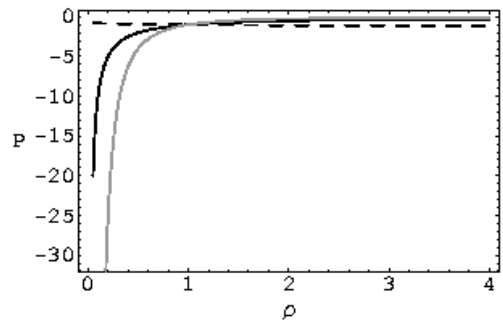}
\caption{The behavior of the pressure  of the generalized
Chaplygin gas is shown in terms of its energy density for negative
values of the parameter $B$. The dashed line corresponds to
$\alpha=-0.1$. The solid darkest (lightest) line corresponds to
$\alpha=1$ ($\alpha=2$). The pressure and energy density has been
redefined as dimensionless quantities given by
$PA^{-1/(1+\alpha)}$ and $\rho A^{-1/(1+\alpha)}$, respectively.
As can be seen, for positive values of $\alpha$ the pressure
reaches extremely negative value when $\rho$ approaches zero.}
\label{chpalyginpressure}
\end{figure}

Once analyzed the late time behaviour of a homogeneous and
isotropic space-time filled by a generalized Chaplygin gas with
the characteristics already mentioned, let us see how it behaves
for smaller scale factors. The energy density vanishes whenever
the scale factor approaches $\bar a$, which can be the case only
for flat and hyperbolic sections. Consequently, at $a=\bar a$ the
pressure may diverge inducing a singularity in the geometry [see
Fig.~\ref{chpalyginpressure}]. Indeed this can be the case if
$\alpha$ is positive. The scalar curvature for $k=\pm 1, 0$ reads
\begin{eqnarray}
R&=&6\left(\dot H +2H^2 +\frac{k}{a^2}\right)\nonumber\\
&=&12D\left[A+\frac{B}{4a^{3(1+\alpha)}}\right]
\left[A+\frac{B}{a^{3(1+\alpha)}}\right]^{-\frac{\alpha}{1+\alpha}},\nonumber\\
\label{curvature}\end{eqnarray}
where the dot represents derivative respect to the cosmic time. As
can be seen, $R$ is well defined at any scale factor for spherical
geometry (we recall $\bar a <a_{\textrm{min}}$). On the other
hand, for $k=-1,0$, the scalar curvature $R$ is finite (even at
$a=\bar a$) as long as $-1<\alpha<0$. The same can de deduced for
positive values of $\alpha$ except at $a=\bar a$, where there is a
divergence of $R$. We would like also to point out that  for flat
sections the FLRW universe presents a bouncing at $a=\bar a$,
which is regular if $-1<\alpha<0$.

\begin{figure}[h]
\includegraphics[width=0.8\columnwidth]{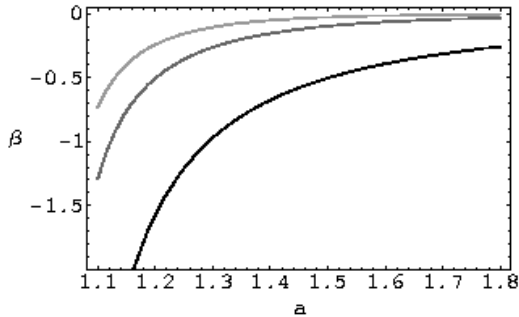}
\caption{This plot shows the behavior of the parameter $\beta$
given in Eq.~(\ref{betagcg}) as a function of the dimensionless
variable $a/\bar a$. The lower graphic (the darkest one)
corresponds to $\alpha=-0.1$. The upper graphic (the lightest one)
corresponds to $\alpha=2$. Finally, the middle graphic shows
$\beta$ for the Chaplygin gas; i.e. $\alpha=1$. As can be see, the
generalized Chaplygin gas approaches a positive cosmological
constant ($\beta=0$) for the largest values of the scale factor.}
\label{betagraph}
\end{figure}

Before concluding this section, we analyze the parameter
$\beta(a)=P/\rho +1$, which somehow quantifies the deviation of
the generalized Chaplygin gas from a cosmological constant [see
Fig.~\ref{betagraph}] and can be expressed in terms of the scale
factor as
\begin{equation}
\beta=\frac{B}{B+Aa^{3(1+\alpha)}}.
\label{betagcg}\end{equation}
As can be expected $\beta$ is negative for the set of parameters
we are considering. At late time, $\beta$ approaches zero; i.e.
the FLRW universe is asymptotically de Sitter. On the other hand,
$\beta$ blows up near $\bar a$ (only for $k=0,-1$). This is
partially due to our oversimplified model. In principle, we should
have considered the other matter contents of the universe, as dark
matter component, which are the dominant components for smaller
scale factors. Indeed, if we consider dark matter (DM) given by
dust, the effective\footnote{The effective value of $\beta$ is
defined as $P/\rho+1$, where now $\rho$ is the total energy
density of the universe and $P$ is the sum of the pressure of the
different material components.} value of $\beta$ approaches the
unity; i.e. the total matter content behaves effectively as dust,
when $a \rightarrow \bar a$, as long as $-1<\alpha<0$. For
positive value of $\alpha$ the effective value of $\beta$ is still
divergent at $\bar a$. This can be understood as a consequence of
the divergence of $P$ near $\bar a$ for positive $\alpha$.

In summary, we have shown that a generalized Chaplygin gas can
describe phantom energy. In addition, this material content can
avoid the occurrence of a cosmic doomsday in the future of the
universe. This is not surprising as the energy density of a
generalized Chaplygin gas approaches a constant positive value for
$B<0$ and $0<\alpha+1$. Consequently, a FLRW universe filled with
this gas is asymptotically de Sitter.

\section{Generalized Chaplygin gas and scalar fields \label{4}}

\begin{figure}[h]
\includegraphics[width=0.8\columnwidth]{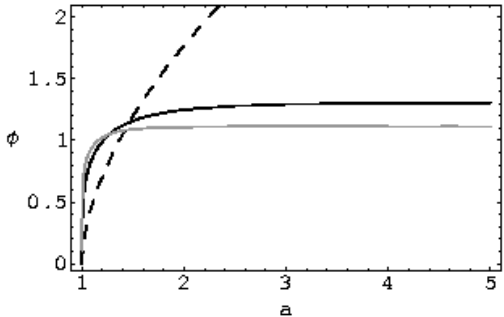}
\caption{This figure shows the behavior of the scalar field $\phi$
as a function of the dimensionless scale factor $a/\bar a$ given
in Eq.~(\ref{phi}). The graphic with dashed line corresponds to
$\alpha=-0.1$, while the graphic with darkest (lightest) full line
corresponds to $\alpha=1$ ($\alpha=2$). In the plot, the scalar
field $\phi$ has been redefined as
$\sqrt{6\pi\textrm{G}}A^{{1}/(2(1+\alpha))}(\phi-\phi_0)$. }
\label{phiplot}
\end{figure}

Up to now, we have described the generalized Chaplygin as a
perfect fluid with a peculiar equation of state (\ref{gcg}). In
the following, we will describe the generalized Chaplygin gas in
terms of scalar fields. First, we will show how the  generalized
Chaplygin gas (with a negative parameter $B$, see
Eq.~(\ref{defB})) can emerge in the context of generalized
Born-Infeld phantom theories. For this purpose, we consider the
Lagrangian $\mathcal{L}_{\phi}$ defined as
\begin{equation}
\mathcal{L}_{\phi}=-A^{\frac{1}{1+\alpha}}\left[1+\left(-g^{\mu\nu}\nabla_\mu\phi\nabla_\nu\phi\right)^{\frac{1+\alpha}{2\alpha}}\right]^{\frac{\alpha}{1+\alpha}},
\label{LGCG}
\end{equation}
where $g_{\mu\nu}$ is the metric of the space-time and $\phi$ is a
scalar field. For a FLRW universe, $\mathcal{L_{\phi}}$ reduces to
\begin{equation}
\mathcal{L}_{\phi}=-A^{\frac{1}{1+\alpha}}\left[1+(\dot\phi)^{\frac{1+\alpha}{\alpha}}\right]^{\frac{\alpha}{1+\alpha}}.
\end{equation}
The dot corresponds to derivative respect to the cosmic time. It
can be shown that the energy density $\rho_\phi$ and the pressure
$P_\phi$ associated to $\mathcal{L_{\phi}}$ reads
\cite{Armendariz-Picon:1999rj},
\begin{eqnarray}
\rho_\phi&=&A^{\frac{1}{1+\alpha}}\left[1+(\dot\phi)^{\frac{1+\alpha}{\alpha}}\right]^{-\frac{1}{1+\alpha}},\nonumber\\
P_\phi&=&-A^{\frac{1}{1+\alpha}}\left[1+(\dot\phi)^{\frac{1+\alpha}{\alpha}}\right]^{\frac{\alpha}{1+\alpha}},
\end{eqnarray}
Consequently, $\rho_\phi$ and $P_\phi$ satisfy a generalized
Chaplygin gas equation of state; i.e.
$P_\phi=-A/\rho_\phi^{\alpha}$. The difference between the
Lagrangian defined by expression (\ref{LGCG}) and the one given in
\cite{chaplygin} is that the kinetic energy term for the scalar
field $\phi$ is negative. In addition, it can be seen that
\begin{equation}
P_\phi/\rho_\phi= -[1+(\dot\phi)^{\frac{1+\alpha}{\alpha}}].
\end{equation}
This expression shows that the scalar field $\phi$ behaves as
phantom energy [see also Eq.~(\ref{4.5})]. Moreover, this
characteristic of $\phi$ allows negative values of the parameter
$B$, defined in Eq.~(\ref{defB}), as has been considered in the
last section. Additionally, on the one hand, the time derivative
of $\phi$ scales with the scale factor as
\begin{equation}
\dot\phi^{\frac{1+\alpha}{\alpha}}=\frac{-B}{B+A a^{3(1+\alpha)}}.
\label{4.5}
\end{equation}
On the other hand, for FLRW universes with flat or hyperbolic
spatial sections filled by a generalized Chaplygin gas with $B<0$
and $-1<\alpha$, the scale factor can takes any value such that
$\bar{a}\leq a$ ($\bar a$ is defined in Eq.~(\ref{bara})).
Consequently, for $-1<\alpha<0$, $\dot\phi$ vanishes when $a$
approaches $\bar{a}$. However, for positive values of $\alpha$,
the time derivative of $\phi$ diverges when $a$ approaches
$\bar{a}$. For large value of the scale factor the opposite
behavior is found; i.e. $\dot\phi$ approaches zero when $0<\alpha$
and diverges when $-1<\alpha<0$. The divergence of $\dot\phi$ is
harmless for large values of the scale factor, as the geometry of
the universe behaves like a de Sitter space-time and,
consequently, there is no singularity. Additionally, in a FLRW
universe with a flat spatial geometry the scalar field $\phi$
varies with the scale factor as
\begin{eqnarray}
&&\phi-\phi_0=\frac{A^{-\frac{1}{2(1+\alpha)}}}{\sqrt{6\pi
G}}\left[1-\left(\frac{\bar a}
{a}\right)^{3(1+\alpha)}\right]^{\frac{1}{2(1+\alpha)}}\nonumber\\
&\times&\textrm{F}\left(\frac{1}{1+\alpha},\frac{1}{2(1+\alpha)};\frac{1}{2(1+\alpha)}+1;1-\left(\frac{\bar
a} {a}\right)^{3(1+\alpha)}\right),\nonumber\\  \label{phi}
\end{eqnarray}
where $-1<\alpha$ and $\phi_0$ is an integration constant
corresponding to the value acquired by $\phi$ at $a=\bar a$. The
scalar field $\phi$ is finite for any value of the scale factor,
whenever $\alpha$ is positive (see footnote \ref{series}). For
$-1<\alpha<0$, the last affirmation remains true except for very
large values of the scale factor, where $\phi$ blows up (see
Fig.~\ref{phiplot}).

\begin{figure}[h]
\includegraphics[width=0.8\columnwidth]{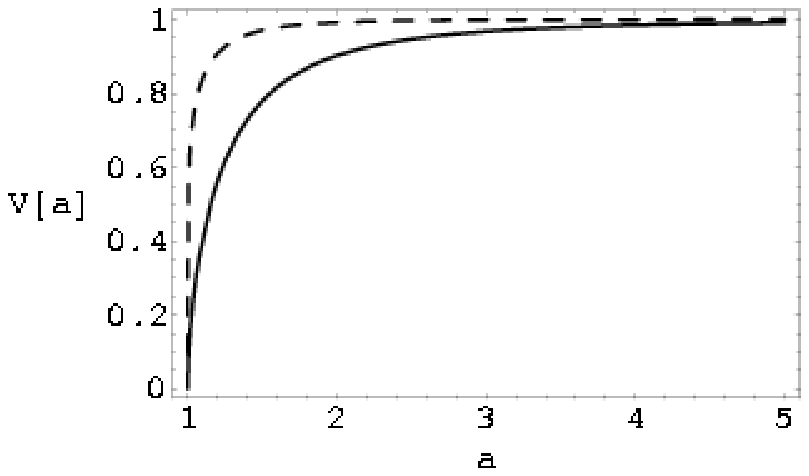}\\\includegraphics[width=0.8\columnwidth]{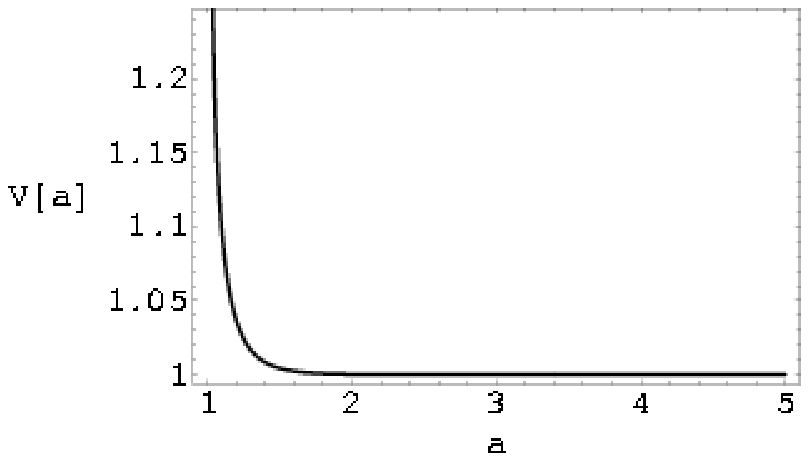}
\caption{These figures show the behavior of $V(a)$ defined in
Eq.~(\ref{vpsi}) as a function of the scale factor. The figure on
the top corresponds to the values $\alpha=-0.1$ (full line) and
$\alpha=1/2$ (dashed line). The figure on the bottom corresponds
to $\alpha=2$. The potential $V(a)$ has been redefined as
$A^{-{1}/(1+\alpha)}V(a)$ and the scale factor as $a/\bar
a$.}\label{vpsiplot}
\end{figure}

In the following, we show how the generalized Chaplygin gas can be
described effectively in terms of a Born-Infeld phantom scalar
field, $\psi$, whose Lagrangian reads
\begin{equation}
\mathcal{L}_{\psi}=-V(\psi)
\sqrt{1-g^{\mu\nu}\nabla_\mu\psi\nabla_\nu\psi}. \label{LBE}
\end{equation}
For a homogenous and isotropic space-time, the energy density
$\rho_{\psi}$ and the pressure $P_{\psi}$ associated to $\psi$
reads \cite{Li,Armendariz-Picon:1999rj}
\begin{equation}
\rho_{\psi}=\frac{V(\psi)}{\sqrt{1+\dot\psi^2}},\quad
P_{\psi}=-V(\psi)\sqrt{1+\dot\psi^2}.
\end{equation}
Obviously, for a Chaplygin gas; i.e. $\alpha=1$, with phantom
energy characteristics, the potential $V(\psi)$ is constant;
$V=\sqrt{A}$ (see Eq.~(\ref{LGCG}) for $\alpha=1$). However, in
general a generalized Chaplygin gas can be described effectively
by a Born-Infeld phantom scalar field, $\psi$,  only when its
potential depends explicitly on $\psi$. In fact, it can be easily
seen that $V$ varies with the scale factor as follows
\begin{equation}
V(a)=\sqrt{A}\left(A+\frac{B}{a^{3(1+\alpha)}}\right)^{\frac{1-\alpha}{2(1+\alpha)}}.
\label{vpsi}\end{equation}
The potential $V$ approaches  a constant value for large values of
the scale factor (we are considering $-1<\alpha$ and $B<0$). In
addition, its behaviour near the minimum value of the scale
factor, $\bar a$, (for $k=0,-1$), depends strongly on the specific
value of the parameter $\alpha$: for $|\alpha|<1$ the potential
vanishes near $\bar a$, for $1<|\alpha|$ the potential blows up
(see Fig.~\ref{vpsiplot}). A main difference between the behaviour
of the scalar fields $\psi$ and $\phi$ is that $\dot\psi$ always
reaches large values near $\bar a$, while this is not necessarily
true for $\dot \phi$. Additionally, it can be seen that for a FLRW
universe with spatially flat sections, the scalar field $\psi$
varies with the scale factor as
\begin{eqnarray}
\psi&-&\psi_0=\frac{A^{-\frac{1}{2(1+\alpha)}}}{\sqrt{6\pi G}
\alpha}\left[1-\left(\frac{\bar a}
{a}\right)^{3(1+\alpha)}\right]^{\frac{\alpha}{2(1+\alpha)}}\nonumber\\
&\times&\textrm{F}\left(\frac12,\frac{\alpha}{2(1+\alpha)};\frac{\alpha}{2(1+\alpha)}+1;1-\left(\frac{\bar
a} {a}\right)^{3(1+\alpha)}\right),\nonumber\\
\label{psi1}\end{eqnarray}
when $\alpha$ is positive. In the last equation $\psi_0$ is a
constant corresponding to the value acquired by $\psi$ at $a=\bar
a $. It can be shown that $\psi$ is finite for any value of the
scale factor (see footnote \ref{series}). On the other hand, for
$-1<\alpha<0$, the scalar field scales with $a$ as
\begin{eqnarray}
\psi-\psi_\infty&=&\frac{A^{-\frac{1}{2(1+\alpha)}}}{\sqrt{6\pi G}
(1+\alpha)}\left(\frac{\bar
a}{a}\right)^{\frac{3(1+\alpha)}{2}}\nonumber\\
&\times&\textrm{F}\left(\frac{2+\alpha}{2(1+\alpha)},\frac12;\frac32;
\left(\frac{\bar a}{a}\right)^{3(1+\alpha)}\right),\,\,
\label{psi2}\end{eqnarray}
where $\psi_\infty$ is the value reached by $\psi$ for very large
scale factors. In this case, the scalar field $\psi$ is well
behaved for any value of $a$, expect at $a=\bar a$ where it blows
up.
\begin{figure}[h]
\includegraphics[width=0.8\columnwidth]{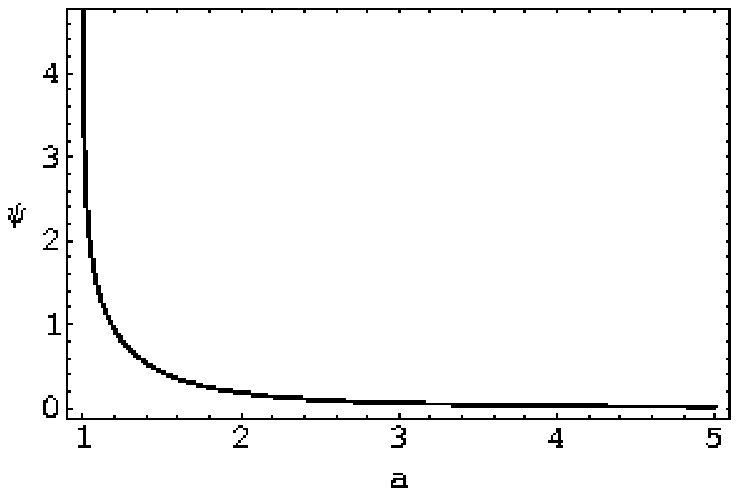}\\
\includegraphics[width=0.8\columnwidth]{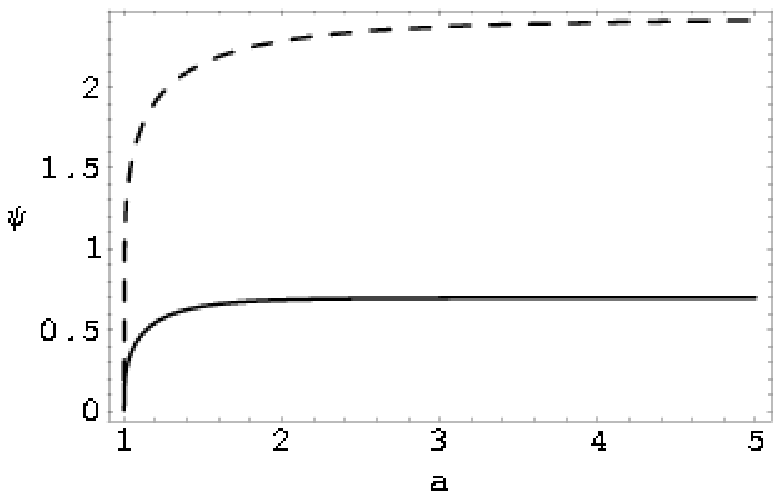}
\caption{These figures show the behavior of $\psi$ defined in
Eqs.~(\ref{psi1}) and (\ref{psi2}) as a function of the scale
factor for a FLRW universe with flat spatial sections. The figure
on the top corresponds to $\alpha=-0.1$. The figure on the bottom
corresponds  to $\alpha=0.5$ (dashed line) and $\alpha=2$ (full
line). In the top plot, the scalar field $\psi$ has been redefined
as $\sqrt{6\pi\textrm{G}}A^{{1}/(2(1+\alpha))}(\psi-\psi_0)$. In
the bottom plot $\psi$ is redefined as
$\sqrt{6\pi\textrm{G}}A^{{1}/(2(1+\alpha))}(\psi-{\psi}_\infty)$.
In both plots the scale factor has been divided by $\bar
a$.}\label{psiplot}
\end{figure}
\begin{figure}[h]
\includegraphics[width=0.7\columnwidth]{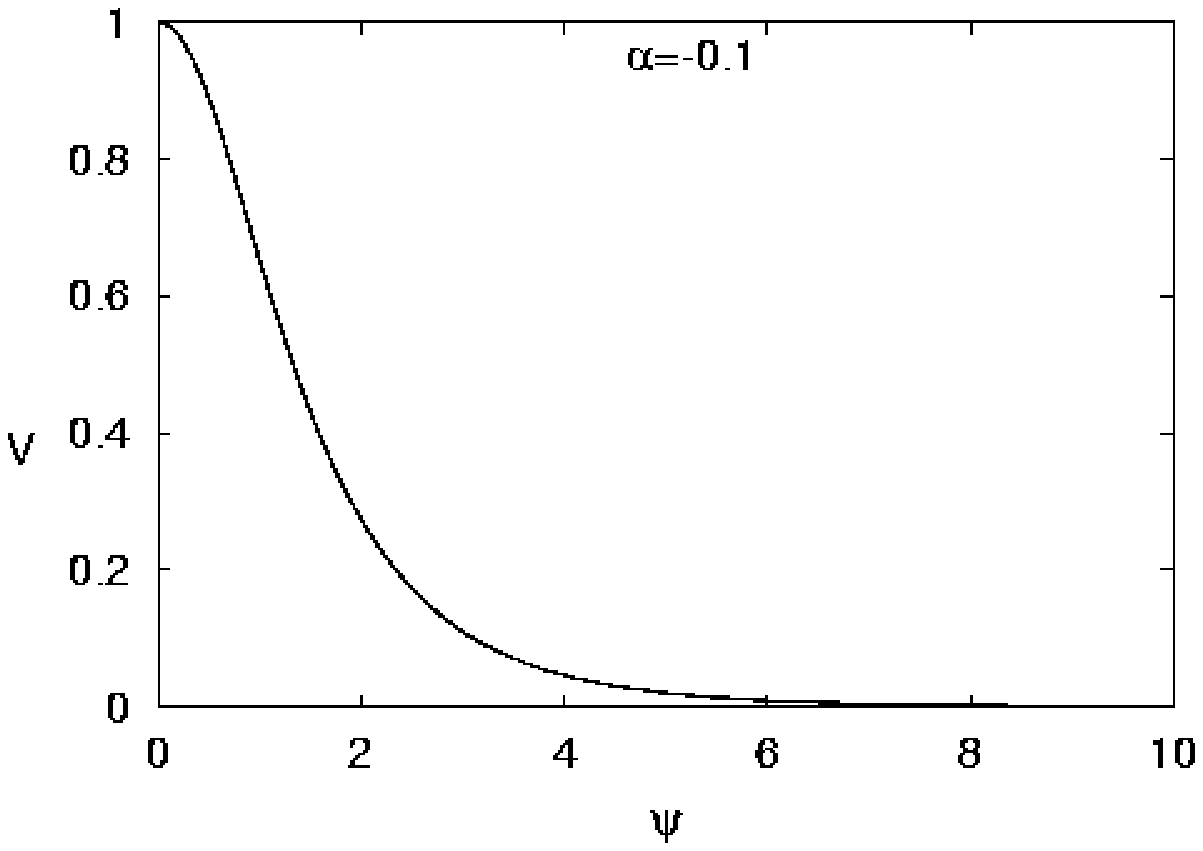}\\
\includegraphics[width=0.7\columnwidth]{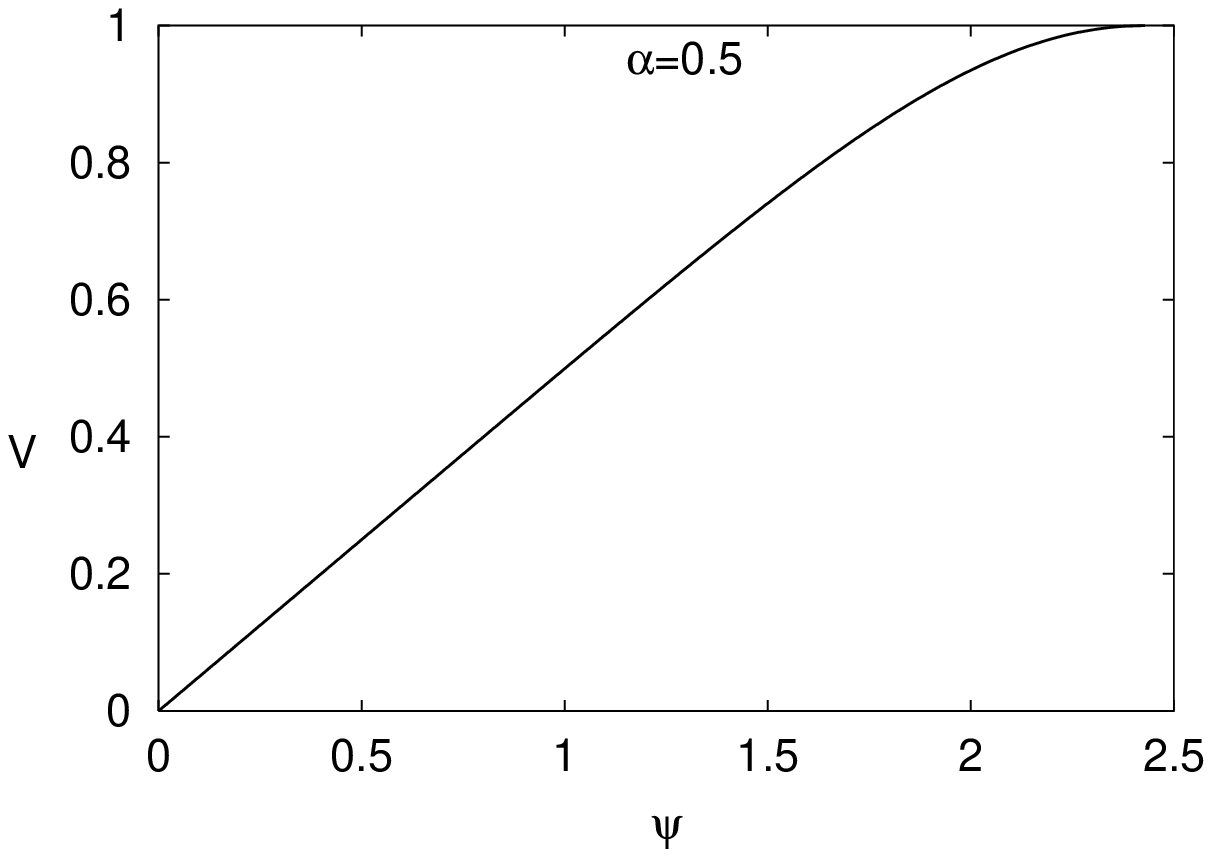}\\
\includegraphics[width=0.7\columnwidth]{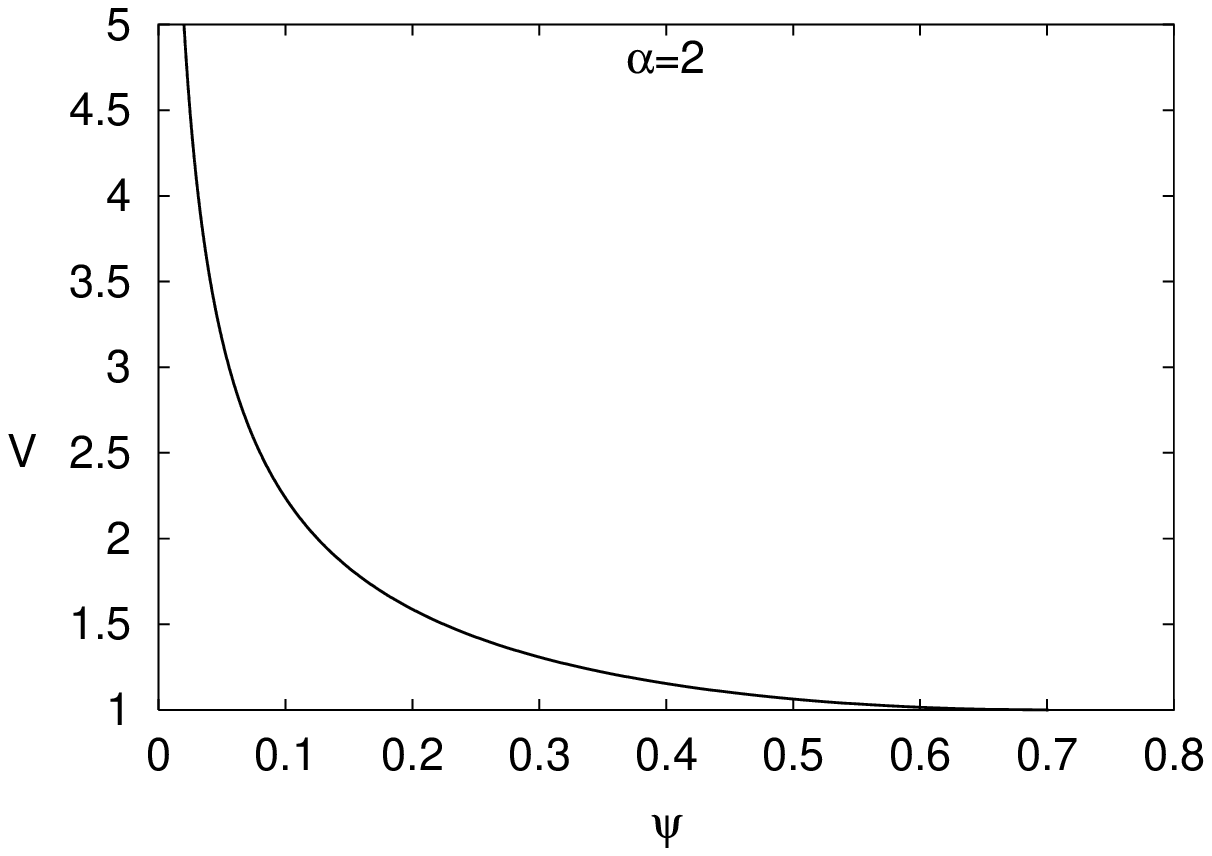}
\caption{These figures show the behavior of $V$ as a function of
the scalar field $\psi$ [see Eqs.~(\ref{vpsi}), (\ref{psi1})and
(\ref{psi2})] for a FLRW universe with flat spatial sections. The
figures starting from the top correspond to $\alpha=-0.1$,
$\alpha=0.5$ and $\alpha=2$, respectively. The potential $V$ has
been redefined as $A^{-{1}/(1+\alpha)}V$. In the top plot $\psi$
is redefined as
$\sqrt{6\pi\textrm{G}}A^{{1}/(2(1+\alpha))}(\psi-{\psi}_\infty)$.
On the other plots, the scalar field $\psi$ has been redefined as
$\sqrt{6\pi\textrm{G}}A^{{1}/(2(1+\alpha))}(\psi-\psi_0)$.}\label{Vpsi-psiplot}
\end{figure}

\begin{figure}[h]
\includegraphics[width=0.8\columnwidth]{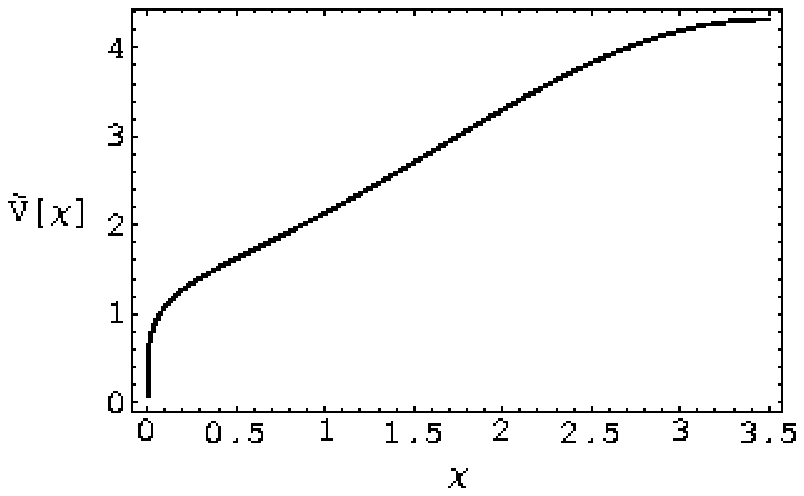}\\
\hspace*{-0.7cm}\includegraphics[width=0.8\columnwidth]{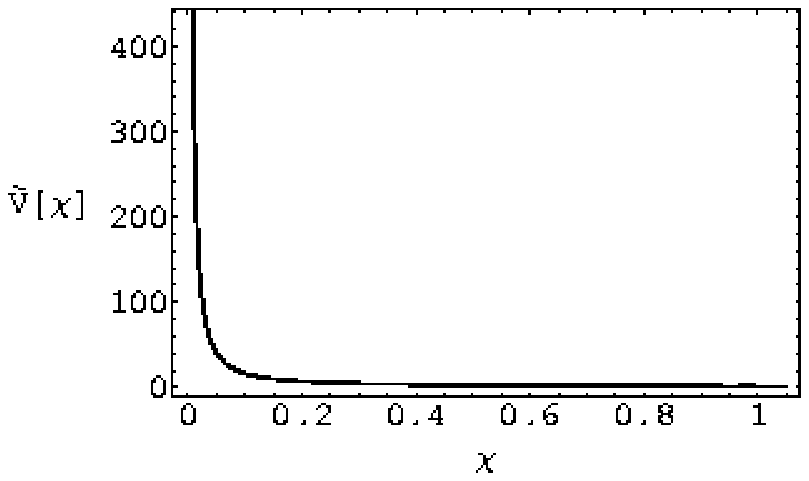}
\caption{These figures show the behaviour of $\tilde V(\chi)$
defined in Eq.~(\ref{Vchi})  as a function of the minimally
coupled scaler field $\chi$ for a FLRW universe with flat spatial
sections. The figure on the top corresponds to $\alpha=-0.1$. The
figure on the bottom corresponds  to $\alpha=2$. In the plots, the
scalar field $\chi$ has been redefined as
$\sqrt{24\pi\textrm{G}}(\chi-\chi_0)$ and the potential as
$2(2/A)^{1/(1+\alpha)}\tilde{V}(\chi)$. As can be seen the
behaviour of $\tilde{V}$ depends on the chosen value of $\alpha$.
For negative values of $\alpha$ the scalar field rolls up the
potential. However, for positive values of $\alpha$, the scalar
field $\chi$ rolls down the potential.}\label{Vchiplot}
\end{figure}

\begin{figure}[h]
\includegraphics[width=0.8\columnwidth]{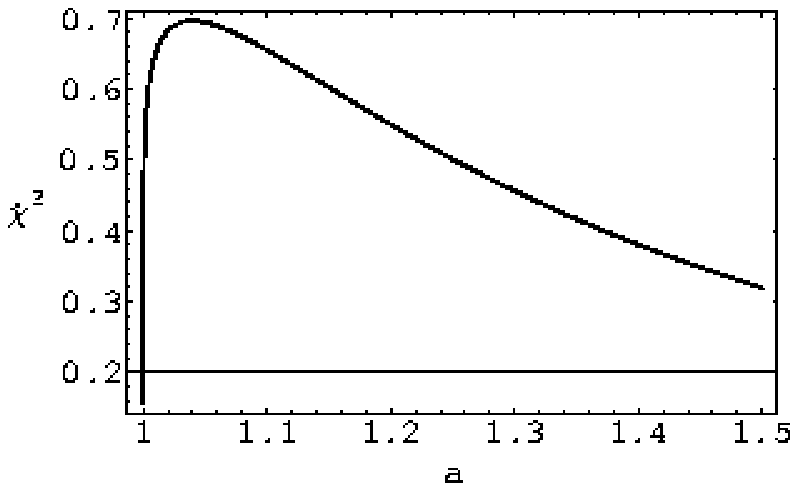} \\
\includegraphics[width=0.8\columnwidth]{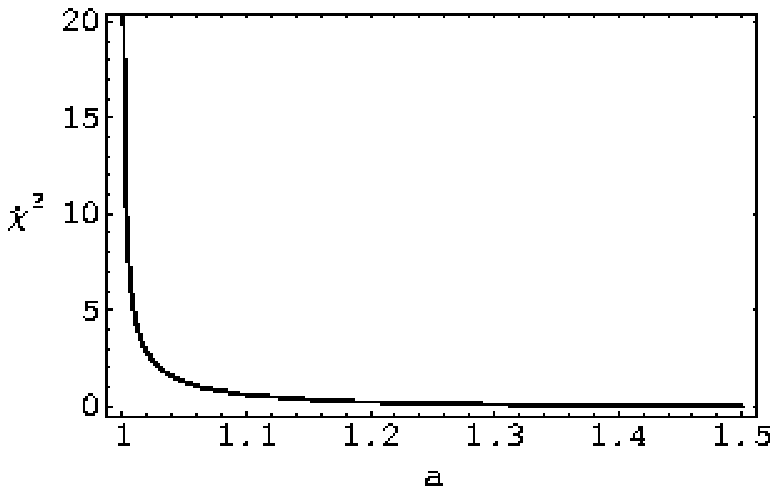}
\caption{The behaviour of the quadratic velocity of the scalar
field $\dot\chi^2$ is shown as a function of $a/\bar{a}$. The
figure on the top corresponds to $\alpha=-0.1$, while the figure
on the bottom corresponds to $\alpha=2$. $\dot\chi^2$ has been
redefined as $A^{-1/(1+\alpha)}\dot\chi^2$.}\label{dotchiplot}
\end{figure}

Finally, we analyze the behaviour of a phantom minimally coupled
scalar field, $\chi$, able to mimic the behaviour of a generalized
Chaplygin gas (for $B$ negative and $-1<\alpha$). In this case,
the energy density $\rho_{\chi}$ and pressure $P_{\chi}$ of the
homogeneous scalar field $\chi$ read
\begin{eqnarray}
\rho_{\chi}=-\frac12{\dot\chi}^2 +\tilde V(\chi), \quad P_{\chi}=
-\frac12{\dot\chi}^2 -\tilde V(\chi).
\end{eqnarray}
If the scalar field $\chi$ simulates  a generalized Chaplygin gas,
the potential $\tilde V$ scales with the scale factor as
\begin{eqnarray}
\tilde{V}(a)=\frac12\left[A+\frac{B}{a^{3(1+\alpha)}}\right]^{-\frac{\alpha}{1+\alpha}}
\left[2A+\frac{B}{a^{3(1+\alpha)}}\right].
\end{eqnarray}
As can be seen, $\tilde{V}(a)$ approaches a constant value when
$a$ blows up. This is not surprising: as we have already mentioned
a FLRW universe filled with a generalized Chaplygin gas is
asymptotically de Sitter. For the scalar field, $\chi$, this
results on $\tilde V(a)$ approaching a non vanishing constant and
a vanishing $\dot\chi$ for very large scale factors. In addition,
$\tilde V(a)$ is finite when $a$ approaches $\bar a$ (for
$k=0,-1$) as long as $-1<\alpha<0$. However, for positive values
of $\alpha$, the potential $\tilde V$ blows up near $\bar a$.
Moreover, If the FLRW universe is spatially flat, $\chi$ is well
behaved for any value of $a$. In fact, its evolution can be
described in term of the scale factor as
\begin{eqnarray}
\chi-\chi_0&=&\frac{1}{(1+\alpha)\sqrt{24\pi\textrm{G}}}\nonumber\\
&\times&\left\{\frac{\pi}{2}- \arcsin\left[2\left(\frac{\bar
a}{a}\right)^{3(1+\alpha)}-1\right]\right\},
\end{eqnarray}
where $\chi_0$ is an integration constant. In a addition, the
potential $\tilde V$ varies with $\chi$ as
\begin{eqnarray}
\tilde
V(\chi)&=&\frac12\left(\frac{A}{2}\right)^{\frac{1}{1+\alpha}}\nonumber\\
&\times&\frac{3-\cos\left[(1+
\alpha)\sqrt{24\pi\textrm{G}}(\chi-\chi_0)\right]}
{\left\{1-\cos\left[(1+
\alpha)\sqrt{24\pi\textrm{G}}(\chi-\chi_0)\right]\right\}^{\frac{\alpha}{1+\alpha}}}.\nonumber\\
\label{Vchi}\end{eqnarray}
The behaviour of the potential $\tilde V(\chi)$ depends strongly
on the value of the parameter $\alpha$ (see Fig.~\ref{Vchiplot}).
For $-1<\alpha<0$ the scalar field rolls up the potential.
However, for positive values of $\alpha$, the scalar field $\chi$
rolls down the potential in contrast with the result obtained in
Ref.~\cite{SSD}. In the first case, the scalar field starts with a
vanishing velocity ($\dot\chi=0$) climbing the potential. Its
velocity continues increasing until it reaches a maximum value and
then it starts decreasing, vanishing for very large values of the
scale factor (see Fig.~\ref{dotchiplot}). In the second case,
$\chi$ starts with an infinite velocity rolling down the
potential. Its velocity is continuously decreasing, until
vanishing when the scale factor blows up (see
Fig.~\ref{dotchiplot}).

\section{A phenomenologically viable model \label{viable}}

\begin{figure}[h]
\includegraphics[width=0.8\columnwidth]{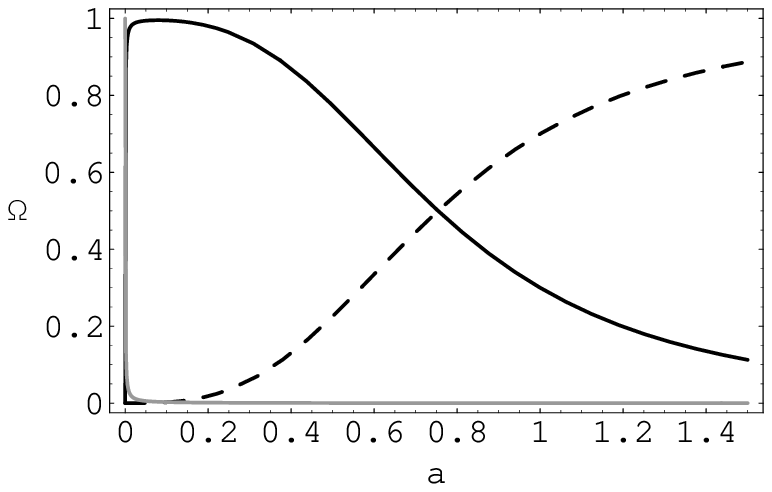}\\
\includegraphics[width=0.8\columnwidth]{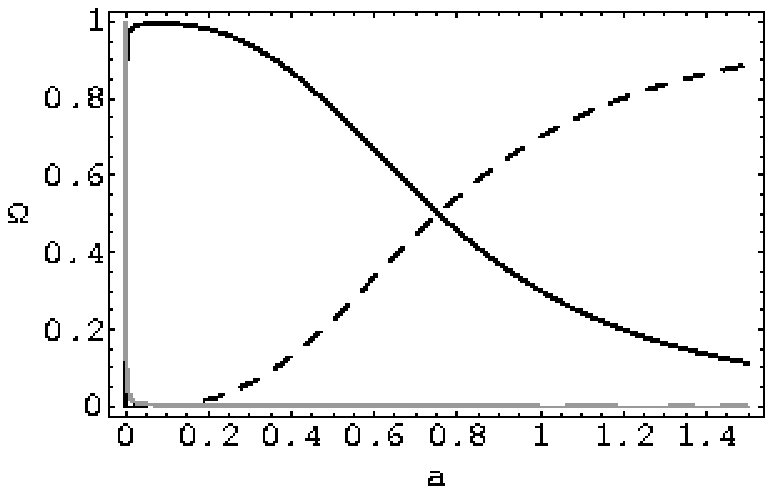}
\caption{These figures shows the behavior of $\Omega_\textrm{DE}$,
$\Omega_\textrm{DM}$ and $\Omega_{R}$ as a function of the scale
factor since the nucleosynthesis epoch; i.e. $\rho_{r}=1 MeV$ .
The lightest graphics correspond to $\Omega_{R}$. The darkest full
line graphics correspond to $\Omega_{\textrm{DM}}$. Finally, the
dashed line graphics correspond to $\Omega_{\textrm{DE}}$. On the
one hand, in the figure of the top, it has been chosen
$\alpha=-0.1$, $A^{1/(1+\alpha)}\simeq 2.8 \times 10^{-11} eV$ and
$|B|^{1/(1+\alpha)}\simeq 3.8 \times 10^{-40} eV$. On the other
hand, on the figure of the bottom, it has been chosen
$\alpha=-0.9$, $A^{1/(1+\alpha)}\simeq 2.8 \times 10^{-10} eV$ and
$|B|^{1/(1+\alpha)}\simeq 1.4 \times 10^{-41} eV$. As can be seen
the behavior is very similar for both choices.} \label{omega}
\end{figure}

In order to study the possible occurrence of a big rip  (in the
future of the universe) caused by phantom energy, it is a good
approximation to consider that the matter content of the universe
is mainly given by phantom energy at very late time (large scale
factors). However, any cosmological viable model able to describe
the actual acceleration of the universe has to take into account
the other material components of the universe, in particular DE.
Consequently, the Friedmann equation reads
\begin{equation}
H^2=\frac{8\pi \textrm G
}{3}\left(\rho_{\textrm{DE}}+\rho_{\textrm{DM}}+\rho_{\textrm{R}}\right),
\end{equation}
where $\rho_{\textrm{DE}}, \rho_{\textrm{DM}}, \rho_{\textrm{R}}$
correspond, respectively, to the energy density of $\textrm{DE}$,
$\textrm{DM}$, and radiation. We will consider that $\textrm{DE}$
is described by a generalized Chaplygin gas with $B<0$ and
$-1<\alpha$ (see Eq.~(\ref{chaplygindensity})) and  the
$\textrm{DM}$ component as a dust fluid. The Friedmann equation
can be rewritten as
\begin{equation}
H^2=H_i^2\left[\frac{\rho_{\textrm{DE}}}{\rho_{C,i}}+\Omega_{\textrm{DM}}\left(\frac{a_i}{a}\right)^3+
\Omega_{\textrm{R}}\left(\frac{a_i}{a}\right)^4\right].
\end{equation}
In the last expression, $H_i, a_i, \rho_{C,i}$ are  the present
Hubble parameter, scale factor and critical energy density. On the
other hand,  $\Omega_{\textrm{DM}}$ and $\Omega_{\textrm{R}}$ are
the density parameters for $\textrm{DM}$  and radiation. The model
can describe the present acceleration of the universe as long as
${\rho_{\textrm{DE}}}/{\rho_{C,i}}\simeq 0.7$,
$\Omega_{\textrm{R}}\simeq 0.3$ and $\Omega_{\textrm{R}}\simeq 0$.
On the other hand, at the radiation dominated era ($T=1 MeV$) the
energy density must be dominated by the $\rho_{\textrm{R}}$ (for a
successful Nucleosynthesis). Considering that the present scale
factor is equal to the unity, the scale factor  in the radiation
dominated era is equal to $2.4\times 10^{-10}$. Our model can
describe a viable cosmological model satisfying all these
requirements for different values of $\alpha$ [see
Fig.~\ref{omega}], although there is a fine tuning of the
parameters $A$ and $B$.

\section{Conclusions \label{conclusion}}

In this paper we study the behaviour of several phantom energy
models able to describe effectively  dark energy. First, we review
the dynamics of a FLRW universe with phantom energy given by a
perfect fluid with a constant equation of state; i.e.
$P=(\beta-1)\rho$ where $\beta$ is constant and negative. It is
shown that the universe hits a big rip independently of the
spatial geometry of the universe. Additionally, it is  shown in
this setup that the presence of a positive constant energy density
in the universe cannot avoid the happening of a cosmic doomsday.
Indeed, the universe hits the big rip sooner than it would be with
$\Lambda=0$.

Secondly, we model dark energy by a phantom generalized Chaplygin
gas; i.e. $P=-A/\rho^{\alpha}$. It is shown that this gas behaves
as a phantom energy as long as the parameter $B$ is negative [see
Eq.~(\ref{defB})]. In addition, the energy density of the gas
increases with the expansion of the universe, vanishing for a
minimum scale factor and approaching a constant value at very late
times (for $-1<\alpha$). Consequently, it can be shown in this
case that the universe will never hit a big rip. In fact, the
universe is asymptotically de Sitter in this case.

The model  involving the phantom generalized Chaplygin gas can
describe the actual acceleration of the universe [see
Sec.~\ref{viable}], where the energy density of the  gas
corresponds  roughly to two thirds of the total energy density of
the universe. However, there is a fine tuning between the
parameters $A$ and $B$ related to the energy density of the gas.

We have also shown how the phantom generalized Chaplygin gas can
appear in a natural way in the context of  phantom  generalized
Born-Infeld theories, where the kinetic energy term for the scalar
field is negative. In addition, we have analyzed different
effective phantom scalar fields which can mimic the behaviour of
this gas. This has been carried out in the context of a
Born-Infeld scalar field and a minimally coupled scalar field.
Although, the dynamics of each of this scalar fields can be quite
different, they provide the same behaviour for the scale factor
and avoid the occurrence of a cosmic doomsday in the future of the
universe.

Finally, we would like to stress that a phantom energy model does
not necessarily imply  a big rip in the future of the universe as
has been shown using a phantom generalized Chaplygin gas, even
without imposing any restriction on the sound speed
\cite{phantom2}. Moreover, a big rip singularity is not
necessarily related to a phantom matter content as has been
recently pointed out in \cite{Barrow:2004xh}. Although there the
big rip singularity happens in a finite cosmic time, scale factor,
energy density and Hubble constant; and the singularity is
associated with a divergence of the pressure.

\acknowledgments

MBL is supported by the Spanish Ministry of Education, Culture and
Sport (MECD). MBL is also partly supported by DGICYT under
Research Project BMF2002 03758. JAJM is supported by the Spanish
MCYT under Research Project BFM2002-00778. The authors thank P.F.
Gonz\'{a}lez-D\'{\i}az and D. Wands for useful discussions. The
authors are also grateful to V. Aldaya, C. Barcelo, M. C. Bertos,
P. Singh, S. Tsujikawa and P. Vargas Moniz for useful comments.
JAJM acknowledges hospitality at the Institute of Cosmology and
Gravitation of the Portsmouth University where part of this work
was carried out.

\appendix*

\section{Explicit expression of $\mathbf{a_{\textrm{min}}}$}

The Friedmann equation (\ref{gcgFriedmann}) for $k=1$ can be
expressed as $H^2=f(a)$ where
\begin{equation}
f(a)=-1/a^2+D\left[A+\frac{B}{a^{3(1+\alpha)}}\right]^{\frac{1}{1+\alpha}}.
\end{equation}
The function $f(a)$ has a unique positive root which we have
denoted $a_{\textrm{min}}$. Its corresponds to the minimum radius
of a FLRW universe filled by a generalized Chaplygin gas with $B$
negative and $-1<\alpha$. The explicit expression of
$a_{\textrm{min}}$  depends on the parameter \cite{X}
\begin{equation}
Q\equiv\frac{B^2}{4A^2}-\frac{D^{-3(1+\alpha)}}{27A^3}.
\end{equation}
For positive value of $Q$; i.e. $(27/4)D^{3(1+\alpha)}AB^2>0$,
$a_{\textrm{min}}$ reads
\begin{equation}
a_{\textrm{min}}=\left[\left(-\frac{B}{2A}+\sqrt{Q}\right)^{\frac13}+
\left(-\frac{B}{2A}-\sqrt{Q}\right)^{\frac13}\right]^{\frac{1}{1+\alpha}}.
\end{equation}
For $Q=0$ that is $B^2=4/(27D^{3(1+\alpha)}A)$, $a_{\textrm{min}}$
is
\begin{equation}
a_{\textrm{min}}=
\frac{1}{\sqrt{D}}\left(\frac{4}{3A}\right)^{\frac{1}{2(1+\alpha)}}.
\end{equation}
Finally, if $(27/4)D^{3(1+\alpha)}AB^2<0$; i.e. negative value of
$Q$, $a_{\textrm{min}}$ can be expressed as
\begin{equation}
a_{\textrm{min}}=\left[2\varrho^{1/3}\cos\left(\frac{\theta}{3}\right)\right]^{\frac{1}{1+\alpha}},
\end{equation}
where
\begin{eqnarray}
\varrho&=&\frac{1}{\sqrt{27D^{3(1+\alpha)}A^3}}, \nonumber\\
\theta&=&\arctan\sqrt{-1+\frac{4}{27D^{3(1+\alpha)}AB^2}}.
\end{eqnarray}
It can be checked that for any value of $Q$, $a_{\textrm{min}}$ is
larger than $\bar a$ given in Eq.~(\ref{bara}).

\end{document}